\documentclass[12pt]{article}
  \usepackage{amsfonts}
  \usepackage{amsmath}
\usepackage{amssymb}
\usepackage{amscd}
\usepackage{graphicx}
\begin{document}
~~
\bigskip
\bigskip
\begin{center}
{\Large {\bf{{{The Henon-Heiles system defined on Lie-algebraically deformed Galilei space-time}}}}}
\end{center}
\bigskip
\bigskip
\bigskip
\begin{center}
{{\large ${\rm {Marcin\;Daszkiewicz}}$ }}
\end{center}
\bigskip
\begin{center}
{ {{{Institute of Theoretical Physics\\ University of Wroc{\l}aw pl.
Maxa Borna 9, 50-206 Wroc{\l}aw, Poland\\ e-mail:
marcin@ift.uni.wroc.pl}}}}
\end{center}
\bigskip
\bigskip
\bigskip
\bigskip
\bigskip
\bigskip
\bigskip
\bigskip
\begin{abstract}
In this article we provide the Henon-Heiles system defined on Lie-algebraically deformed nonrelativistic space-time with the commutator of two spatial directions
proportional to time. Particularly,
we demonstrate that in such a model the total energy is not conserved and for this reason the role of control parameter is taken by the initial energy  value
$E_{0,{\rm tot}} = E_{{\rm tot}}(t=0)$. Besides, we show that in contrast with
 the commutative case, for chosen  values of deformation parameter $\kappa$, there appears chaos in the system
for initial total energies $E_{0,{\rm tot}}$  below the threshold  $E_{0,{\rm th}} = 1/6$.
\end{abstract}
\bigskip
\bigskip
\bigskip
\bigskip
\bigskip
\bigskip
\bigskip
\bigskip
\bigskip
 \eject
\section{{{Introduction}}}

At the beginning of 70's Edward Lorenz proposed his widely-known ''model of weather'' containing the system of
nonlinear and strongly sensitive with respect initial conditions differential equations \cite{lorenc}. Since this time there appears
a lot of papers dealing with classical and quantum systems of which dynamic remains chaotic; the most popular of them are:
Henon-Heiles system \cite{henon}, Rayleigh-Bernard system \cite{rbsystem}, jerk equation \cite{jerk},
Duffing equation \cite{duffing}, double pendulum \cite{dpendulum}, \cite{dpendulum1},
pendulum with  forced damping \cite{dpendulum}, \cite{dpendulum1} and
quantum forced damped oscillator model \cite{quantumdamped}. The first of them, so-called Henon-Heiles system, has been provided in
pure astrophysical context, i.e., it concerns the problem of nonlinear motion of a star around of a galactic center where the motion is
restricted to a plane. It is defined by the following Hamiltonian function
\begin{eqnarray}
H(p,x) \;=\; \frac{1}{2}\sum_{i=1}^2\; \left(p_i^2 + x_i^2\right) + x_1^2x_2 - \frac{1}{3}x_2^3\;,\label{ham}
\end{eqnarray}
which in cartesian coordinates $x_1$ and $x_2$ describes the set of two nonlinearly coupled harmonic oscillators.
In polar coordinates $r$ and $\theta$ it corresponds to the particle moving in noncentral potential of the form
\begin{eqnarray}
V(r,\varphi) \;=\; \frac{r^2}{2} + \frac{r^3}{3}\sin\left(3\varphi\right)\;,\label{pot}
\end{eqnarray}
with $x_1 = r\cos\varphi$ and $x_2 = r\sin\varphi$. The above model has been inspired by observations indicating that star moving
in a weakly  perturbated central potential should has apart of  total energy $E_{\rm{tot}}$ constant in time also the second
conserved physical quantity $I$. It has been demonstrated with use of so-called
Poincare section method, that such a situation appears in the case of Henon-Heiles system only for the values of control parameter $E_{\rm{tot}}$ below
the threshold $E_{\rm th} = 1/6$. For higher energies the trajectories in phase space become chaotic and the quantity $I$ does not exist (see e.g. \cite{tabor}, \cite{genhamhh}).

Our aim in this paper is to study further the impact of noncommutative generalization of the  model on its chaotic properties.
Recently, there has been proposed in article \cite{henoncanondef} the noncommutative counterpart of Henon-Heiles system defined on the following
canonically deformed Galilei space-time \cite{oeckl}-\cite{dasz1}\footnote{The Lie-algebraically noncommutative space-time has been  defined as the quantum
representation spaces, so-called Hopf modules (see e.g. \cite{oeckl}, \cite{chi}), for the canonically deformed quantum Galilei Hopf algebras $\,{\mathcal U}_\kappa({ \cal G})$.},\footnote{It should be noted that in accordance with the
Hopf-algebraic classification  of all deformations of relativistic
and nonrelativistic symmetries (see references \cite{class1}, \cite{class2}),
apart of canonical \cite{oeckl}-\cite{dasz1} space-time noncommutativity, there also exist  Lie-algebraic \cite{dasz1}-\cite{lie1} and quadratic \cite{dasz1}, \cite{lie1}-\cite{paolo} type of quantum spaces.}
\begin{equation}
[\;t,\hat{x}_{i}\;] = 0\;\;\;,\;\;\; [\;\hat{x}_{i},\hat{x}_{j}\;] = 
i\theta_{ij}
\;,\label{nhspace}
\end{equation}
with $\theta_{ij}=-\theta_{ji} = {\rm const.}$ Particularly, there has been provided the Hamiltonian function of the model as well as the corresponding
canonical equations of motion. Besides, it has been demonstrated that for proper values of deformation parameter $\theta$ and for proper values of control parameter
$E_{\rm{tot}}$ there appears (much more intensively) chaos in the system. Consequently, in such a way it has been shown  the impact of most simple space-time noncommutativity
on the basic dynamical properties of this important classical chaotic model. It should be noted, however,  that there exist a lot of papers dealing with noncommutative classical and quantum  mechanics (see e.g.
\cite{mech}-\cite{qmgnat}) as well as with field theoretical systems (see e.g. \cite{prefield}-\cite{fiorewess}), in which  the quantum
space-time is not classical. Such  models follow, in particular, from  formal  arguments based mainly  on
Quantum Gravity \cite{2}, \cite{2a} and String Theory 
\cite{recent}, \cite{string1} indicating that space-time at Planck
scale  becomes noncommutative.

In present article we provide another noncommutative Henon-Heiles system defined on the following  Lie-algebraically deformed space-time  \cite{dasz1}
\begin{equation}
[\;t,\hat{x}_{i}\;] = 0\;\;\;,\;\;\; [\;\hat{x}_{i},\hat{x}_{j}\;] =
\frac{i}{\kappa}t \epsilon_{ij}
\;,\label{liespace}
\end{equation}
with deformation parameter $\kappa$. In such a way we get the model with time-dependent Hamiltonian function in which the role of control parameter is taken over by the initial value of total energy $E_{0,\rm{tot}}= E_{\rm{tot}}(t=0)$. We shall demonstrate with use of Poincare section method, that contrary to the commutative case, there appears for suitable values of deformation parameter $\kappa$  chaos below the threshold
$E_{0,\rm{th}}=1/6$. We interpret such a result as an effect of the time dependence of  Hamiltonian
function of the model.

The paper is organized as follows. In second Section we briefly remind the basic properties of classical Henon-Heiles system. In Section 3 we recall
Lie-algebraic noncommutative quantum Galilei space-time proposed in article \cite{dasz1}. Section 4 is devoted to the new Lie-algebraically deformed Henon-Heiles model while the conclusions are discussed in Sect. 5, describing final remarks.

\section{{{Classical Henon-Heiles model}}}

As it was already mentioned the Henon-Heiles system is defined by the following Hamiltonian function
\begin{eqnarray}
H(p,x) \;=\; \frac{1}{2}\sum_{i=1}^2\; \left(p_i^2 + x_i^2\right) + x_1^2x_2 - \frac{1}{3}x_2^3\;,\label{ham1}
\end{eqnarray}
with canonical variables $(p_i,x_i)$ satisfying
\begin{equation}
\{\;x_i,x_j\;\} = 0 =\{\;p_i,p_j\;\}\;\;\;,\;\;\; \{\;x_i,p_j\;\}
=\delta_{ij}\;, \label{classpoisson}
\end{equation}
i.e., it describes the system of two nonlineary coupled one-dimensional harmonic oscillator models. One can check that the corresponding
canonical equations of motion take the form
\begin{eqnarray}
&& \dot{p}_1\;=\; -\frac{\partial H}{\partial x_1}\;=\; -x_1 -2x_1x_2\;\;\;,\;\;\;
\dot{x}_i\;=\; \frac{\partial H}{\partial p_i}\;=\; p_i\;,\label{eqs2}\\[5pt]
&&\dot{p}_2\;=\; -\frac{\partial H}{\partial x_2}\;=\; -x_2 -x_1^2 + x_2^2 \;,
\label{eqs3}
\end{eqnarray}
while the proper Newton equations look as follows
\begin{equation}
\left\{\begin{array}{rcl}
\ddot{x}_1&=& -x_1 - 2x_1x_2\\[5pt]
\ddot{x}_2&=&  -x_2 -x_1^2 + x_2^2 \;.
\end{array}\right.\label{neqs2}
\end{equation}
Besides, it is easy to see that the conserved in time total energy of the model is given by
\begin{eqnarray}
E_{\rm tot} \;=\; \frac{1}{2}\sum_{i=1}^2\; \left(\dot{x}_i^2 + x_i^2\right) + x_1^2x_2 - \frac{1}{3}x_2^3\;.\label{energy}
\end{eqnarray}

In order to illustrate the basic properties of discussed system we find numerically the
Poincare maps in two dimensional phase space $(x_2,p_2)$ for section $x_1 = 0$ and for four fixed values of total energy $E_{\rm{tot}}$ below the threshold $E_{\rm th} = 1/6$: $E_{\rm tot} = 0.00000015$, $E_{\rm tot} = 0.000015$, $E_{\rm tot} = 0.0015125$ and $E_{\rm tot} = 0.03125$; the obtained results are summarized on {\bf Figures 1 - 4} respectively. Consequently, one can see that in such a case the all trajectories (in fact) remain completely regular\footnote{The calculations are performed for single trajectory with initial condition $x_1(0) = \left({2}{E_{\rm tot}}\right)^{\frac{1}{2}}$ and $x_2(0) = p_1(0) = p_2(0) = 0$.}.

\section{Lie-algebraically deformed Galilei space-time}

In this section we very shortly recall the basic facts associated with the (twisted) canonically deformed  Galilei Hopf algebra
$\;{\cal U}_{\kappa}({\cal G})$ and with the
corresponding quantum space-time \cite{dasz1}.  First of all, it should be noted that in accordance with Drinfeld  twist procedure \cite{drin} the algebraic sector of
Hopf structure $\;{\cal U}_{\kappa}({\cal G})$ remains
undeformed
\begin{eqnarray}
&&\left[\, K_{ij},K_{kl}\,\right] =i\left( \delta
_{il}\,K_{jk}-\delta
_{jl}\,K_{ik}+\delta _{jk}K_{il}-\delta _{ik}K_{jl}\right) \;,  \label{ff} \\
&~~&  \cr &&\left[\, K_{ij},V_{k}\,\right] =i\left( \delta
_{jk}\,V_i-\delta _{ik}\,V_j\right)\;\; \;, \;\;\;\left[
\,K_{ij},\Pi_{\rho }\right] =i\left( \eta _{j \rho }\,\Pi_{i }-\eta
_{i\rho }\,\Pi_{j }\right) \;, \label{nnn}
\\
&~~&  \cr &&\left[ \,V_i,V_j\,\right] = \left[ \,V_i,\Pi_{j
}\,\right] =0\;\;\;,\;\;\;\left[ \,V_i,\Pi_{0 }\,\right]
=-i\Pi_i\;\;\;,\;\;\;\left[ \,\Pi_{\rho },\Pi_{\sigma }\,\right] =
0\;,\label{ff1}
\end{eqnarray}
where $K_{ij}$, $\Pi_0$, $\Pi_i$ and $V_i$ can be identified with  rotation, time translation, momentum and boost operators respectively. Besides,
the coproducts and antipodes of such  algebra take the form\footnote{Indices $k$ and $l$ are fixed and $a \wedge b = a\otimes b- b\otimes a$.}
\begin{eqnarray}
 \Delta_{{\kappa}}(\Pi_0)&=&\Delta _0(\Pi_0) +
\frac{1}{2{{\kappa}}} \Pi_l \wedge \Pi_k\;,\label{coppy1}\\
 &~~&  \cr
\Delta_{{\kappa}}(\Pi_i)&=&\Delta
_0(\Pi_i)\;\;\;,\;\;\;\Delta_{{\kappa}}(V_i)=\Delta_0(V_i)\;,\label{cop0}\\
 &~~&  \cr
\Delta_{{\kappa}}(K_{ij})&=&\Delta_0(K_{ij})+
\frac{i}{2{{\kappa}}}\left[\,K_{ij},V_k\,\right]\wedge \Pi_l \;+\label{coppy100}\\
 &~~&  \cr
&~~&~~~~~~~~~~~~~~~~~~~~~~~~~~+\;\frac{1}{2{{\kappa}}}V_k \wedge(\delta_{il}\Pi_j
-\delta_{jl}\Pi_i)  \;, \nonumber
\end{eqnarray}
while the corresponding quantum space-time can be defined as the representation space, so-called Hopf modules (see e.g. \cite{oeckl}, \cite{chi}), for the Lie-algebraically deformed Hopf structure
$\;{\cal U}_{\kappa}({\cal G})$; it looks as follows
\begin{equation}
[\;t,\hat{x}_{i}\;] = 0\;\;\;,\;\;\; [\;\hat{x}_{i},\hat{x}_{j}\;] =
i\frac{t}{\kappa}(\delta _{l i }\delta_{k j }-\delta _{
ki }\delta _{l j
})
\;,\label{liespace1}
\end{equation}
and for  deformation parameter $\kappa$  approaching  infinity it becomes commutative.

\section{{{Lie-algebraically deformed classical Henon-Heiles system}}}

Let us now turn to the Henon-Heiles model defined on quantum space-time (\ref{liespace}).
In first step of our construction we extend the Lie-algebraically deformed space to the whole algebra of momentum
and position operators as follows
(see e.g. \cite{chaihydro2}-\cite{romero1})\footnote{The correspondence relations are $\{\;\cdot,\cdot\;\} = \frac{1}{i}\left[\;\cdot,\cdot\;\right]$.}
\begin{eqnarray}
&&\{\;\hat{ x}_{1},\hat{ x}_{2}\;\} = \frac{2t}{\kappa}\;\;\;,\;\;\;
\{\;\hat{ p}_{i},\hat{ p}_{j}\;\} = 0\;\;\;,\;\;\; \{\;\hat{ x}_{i},\hat{ p}_{j}\;\} = \delta_{ij}\;.\label{rel1}
\end{eqnarray}
 One can check that relations
(\ref{rel1}) satisfy the Jacobi identity and for deformation parameter
$\kappa$ approaching infinity become classical. \\
Next, by analogy to the commutative case we define the corresponding Hamiltonian function by\footnote{Such a construction of deformed Hamiltonian
function (by replacing the commutative variables $(x_i,p_i)$ by noncommutative ones $({\hat x}_i, {\hat p}_i)$) is well-known in the
literature - see e.g. \cite{chaihydro1}, \cite{chaihydro2} and \cite{exem}.}
\begin{eqnarray}
H(\hat{p},\hat{x}) \;=\; \frac{1}{2}\sum_{i=1}^2\; \left(\hat{p}_i^2 + \hat{x}_i^2\right) + \hat{x}_1^2\hat{x}_2 - \frac{1}{3}\hat{x}_2^3\;,\label{nonham}
\end{eqnarray}
with the
noncommutative operators $({\hat x}_i, {\hat p}_i)$  represented by the classical
ones $({ x}_i, { p}_i)$ as  \cite{romero1}-\cite{kijanka}
\begin{eqnarray}
{\hat x}_{1} &=& { x}_{1} - { \frac{t}{\kappa}}p_2\;,\label{rep1}\\[5pt]
{\hat x}_{2} &=& { x}_{2} +{\frac{t}{\kappa}}p_1
\;,\label{rep2}\\[5pt]
{\hat p}_{1} &=& { p}_{1}\;\;,\;\;{\hat p}_{2} \;=\; { p}_{2}\;.\label{rep2a}
\end{eqnarray}
Consequently, we have
\begin{eqnarray}
H({p},{x},t) &=&
\frac{1}{2M\left(\frac{t}{\kappa}\right)}\left({{{p}}_1^2}+{{{p}}_2^2} \right)  +
\frac{1}{2}M\left(\frac{t}{\kappa}\right)\Omega^2\left(\frac{t}{\kappa}\right)\left({{{x}}_1^2}+{{{x}}_2^2} \right)
- S\left(\frac{t}{\kappa}\right)L\;+ \label{2dh1}\\
&+& \left(x_1 - {\frac{t}{\kappa}p_2}\right)^2 \left(x_2 + {\frac{t}{\kappa}p_1}\right)
- \frac{1}{3}\left(x_2 + {\frac{t}{\kappa}p_1}\right)^3\;,\nonumber
\end{eqnarray}
where
\begin{eqnarray}
&&L = x_1p_2 - x_2p_1\;, \\[5pt]
&&\frac{1}{M\left(\frac{t}{\kappa}\right)} = 1 +\left(\frac{t}{\kappa}\right)^2 \;,\\[5pt]
&&\Omega\left(\frac{t}{\kappa}\right) = \sqrt{\left(1
+\left(\frac{t}{\kappa}\right)^2 \right)}\;,
\end{eqnarray}
and
\begin{eqnarray}
S\left(\frac{t}{\kappa}\right)=\frac{t}{\kappa}\;.
\end{eqnarray}
We see that the Hamiltonian function depends on time and for this reason the role of control parameter in the model is taken by initial value of total energy $E_{0,{\rm tot}}=E_{\rm tot}(t=0)$.
Besides, using the
formula (\ref{2dh1}) one gets the following canonical  equations
of motion\\
\begin{equation}
\left\{\begin{array}{rcl}
\dot{x}_1 &=& \frac{1}{M\left(\frac{t}{\kappa}\right)}p_1 + S\left(\frac{t}{\kappa}\right)x_2 +
\left[\left(x_1- \frac{t}{\kappa}p_2\right)^2 - \left(x_2+ \frac{t}{\kappa}p_1\right)^2\right]\frac{t}{\kappa}\;\\[5pt]
&~~&~\cr
\dot{x}_2 &=& \frac{1}{M\left(\frac{t}{\kappa}\right)}p_2 - S\left(\frac{t}{\kappa}\right)x_1 - 2\left[x_2+ \frac{t}{\kappa}p_1\right]
\left[x_1- \frac{t}{\kappa}p_2\right]\frac{t}{\kappa}\;\\[5pt]
&~~&~\cr
\dot{p}_1 &=& -{M\left(\frac{t}{\kappa}\right)}\Omega^2\left(\frac{t}{\kappa}\right)x_1 + S\left(\frac{t}{\kappa}\right)p_2 -
2\left[x_2+ \frac{t}{\kappa}p_1\right]\left[x_1- \frac{t}{\kappa}p_2\right]\;\\[5pt]
&~~&~\cr
\dot{p}_2 &=& -{M\left(\frac{t}{\kappa}\right)}\Omega^2\left(\frac{t}{\kappa}\right)x_2 - S\left(\frac{t}{\kappa}\right)p_1 - \left[x_1- \frac{t}{\kappa}p_2\right]^2 +
\left[x_2+ \frac{t}{\kappa}p_2\right]^2\;,
\label{xxxlie2newton1}\end{array}\right.
\end{equation}\\
which for deformation parameter running to infinity become classical.

Similarly to the undeformed case we
find numerically the
Poincare maps in two dimensional phase space $(x_2,p_2)$ for section $x_1=0$.  We perform our calculations for exactly the same
initial values of total energy as in commutative model,
however, this time, apart of parameter
$E_{0,{\rm tot}}$ we take under consideration the parameter of deformation $\kappa$. Consequently, we derive the Poincare sections of phase space
for four pairs:
$\left(E_{0,{\rm tot}}=0.00000015, 1/\kappa =0.0001\right)$,  $\left(E_{0,{\rm tot}}=0.000015, 1/\kappa =0.001\right)$,  $(E_{0,{\rm tot}}=0.0015125,$ $1/\kappa =0.001)$ and
$\left(E_{0,{\rm tot}}=0.03125,
1/\kappa =0.0001\right)$, and we represent them on {\bf Figures 5 - 8} respectively\footnote{As in the commutative case the calculations are performed for single trajectory with initial condition $x_1(0) = \left({2}{E_{0,{\rm tot}}}\right)^{\frac{1}{2}}$ and $x_2(0) = p_1(0) = p_2(0) = 0$.}. In such a way we demonstrate
that contrary to the undeformed Henon-Heiles system there appears chaos in the model for initial energies $E_{0,\rm{tot}}$ below the threshold $E_{0,{\rm th}}=1/6$.

\section{{{Final remarks}}}

In this article we provide the Henon-Heiles system defined on Lie-algebraically deformed nonrelativistic space-time with the commutator of two spatial directions
proportional to time. Particularly,
we demonstrate that in such a model the total energy is not conserved and for this reason the role of control parameter is taken by the initial energy  value
$E_{0,{\rm tot}} = E_{{\rm tot}}(t=0)$. Besides, we show that in contrast with
 the commutative case, for chosen  values of deformation parameter $\kappa$, there appears chaos in the system
for initial total energies $E_{0,{\rm tot}}$  below the threshold  $E_{0,{\rm th}} = 1/6$.

It should be noted that the present studies can be extended in various ways. First of all one may consider more complicated noncommutative Henon-Heiles models defined, for example, on the quadratic space-times provided in article \cite{dasz1}. Besides, it is possible to investigate the Lie-algebraic  deformation of so-called generalized Henon-Heiles systems given by the following Hamiltonian function
\begin{eqnarray}
H(p,x) \;=\;  \frac{1}{2} \left(p_1^2 + p_2^2\right) + \delta x_1^2 + (\delta+\Omega)x_2^2 + \alpha x_1^2x_2 +\alpha\beta x_2^3\;,\label{genhamh}
\end{eqnarray}
with arbitrary coefficients $\alpha$, $\beta$, $\delta$ and $\Omega$ respectively. It should be noted that the properties of commutative models described by function (\ref{genhamh})
are quite interesting. For example, it is well-known (see e.g. \cite{p1}-\cite{p10} and references therein) that such systems remain integrable only in the Sawada-Kotera case: with $\beta =1/3$ and
$\Omega = 0$, in the KdV case: with $\beta =2$ and arbitrary  $\Omega$ as well as in the Kaup-Kupershmidt case: with $\beta = 16/3$ and $\Omega =15\delta$. Besides, there has been provided in articles \cite{bal1} and \cite{bal2}\footnote{See also references therein.} the different types of integrable perturbations of mentioned above (integrable) models such as, for example, $q^{-2}$ perturbations, the Ramani series of polynomial deformations and the rational perturbations.  Consequently, the impact of the Lie-algebraic deformation (\ref{liespace}) on the above dynamical structures (in fact) seems to be very interesting. For this reason
the works in this direction already started and are in progress.

\section*{Acknowledgments}
The author would like to thank J. Lukierski
for valuable discussions.\\
This paper has been financially supported by Polish Ministry of
Science and Higher Education grant NN202318534.

\eject

$~~~~~~~~~~~~~~~~~~$
\\
\\
\\
\\
\\
\begin{figure}[htp]
\includegraphics[width=\textwidth]{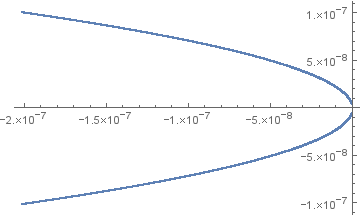}
\caption{Poincare map in two dimensional phase space $(x_2,p_2)$ for section $x_1 = 0$ and for total energy $E_{\rm tot} = 0.00000015$ lying down below the threshold $E_{\rm th} = 1/6$. Trajectory is completely regular - there is no chaos in the system.}\label{rysunek1}
\end{figure}
\eject

$~~~~~~~~~~~~~~~~~~$
\\
\\
\\
\\
\\
\begin{figure}[htp]
\includegraphics[width=\textwidth]{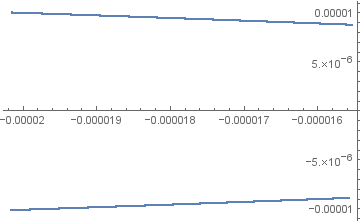}
\caption{Poincare map in two dimensional phase space $(x_2,p_2)$ for section $x_1 = 0$ and for total energy $E_{\rm tot} = 0.000015$  lying down below the threshold $E_{\rm th} = 1/6$. Trajectory is completely regular - there is no chaos in the system.}\label{rysunek2}
\end{figure}
\eject

$~~~~~~~~~~~~~~~~~~$
\\
\\
\\
\\
\\
\begin{figure}[htp]
\includegraphics[width=\textwidth]{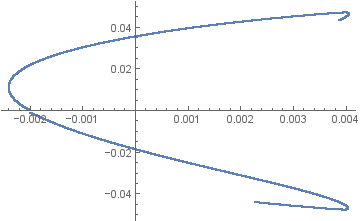}
\caption{Poincare map in two dimensional phase space $(x_2,p_2)$ for section $x_1 = 0$ and for total energy $E_{\rm tot} = 0.0015125$ lying down below the threshold $E_{\rm th} = 1/6$. Trajectory is completely regular - there is no chaos in the system.}\label{rysunek3}
\end{figure}

\eject

$~~~~~~~~~~~~~~~~~~$
\\
\\
\\
\\
\\
\begin{figure}[htp]
\includegraphics[width=\textwidth]{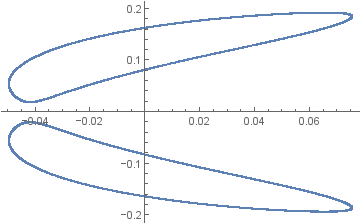}
\caption{Poincare map in two dimensional phase space $(x_2,p_2)$ for section $x_1 = 0$ and for total energy $E_{\rm tot} = 0.03125$ lying down below the threshold $E_{\rm th} = 1/6$. Trajectory is completely regular - there is no chaos in the system.}\label{rysunek3}
\end{figure}

\eject

$~~~~~~~~~~~~~~~~~~$
\\
\\
\\
\\
\\
\begin{figure}[htp]
\includegraphics[width=\textwidth]{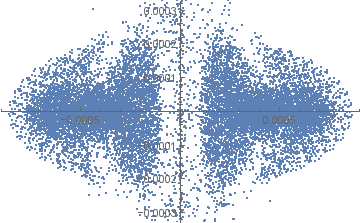}
\caption{Poincare map in two dimensional phase space $(x_2,p_2)$ for section $x_1 = 0$ and for deformation parameter $1/\kappa=0.0001$ as well as for initial value of total energy $E_{0,{\rm tot}} = 0.00000015$ lying down below the threshold $E_{0,{\rm th}} = 1/6$. Trajectory is partially chaotic.}\label{rysunek3}
\end{figure}

\eject

$~~~~~~~~~~~~~~~~~~$
\\
\\
\\
\\
\\
\begin{figure}[htp]
\includegraphics[width=\textwidth]{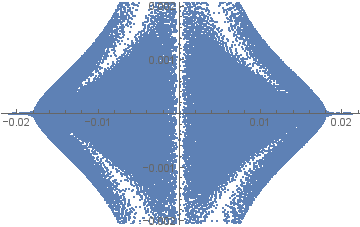}
\caption{Poincare map in two dimensional phase space $(x_2,p_2)$ for section $x_1 = 0$ and for deformation parameter $1/\kappa=0.001$ as well as for initial value of total energy $E_{0,{\rm tot}} = 0.000015$ lying down below the threshold $E_{0,{\rm th}} = 1/6$. Trajectory is almost completely  chaotic.}\label{rysunek3}
\end{figure}

\eject

$~~~~~~~~~~~~~~~~~~$
\\
\\
\\
\\
\\
\begin{figure}[htp]
\includegraphics[width=\textwidth]{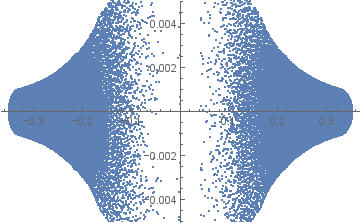}
\caption{Poincare map in two dimensional phase space $(x_2,p_2)$ for section $x_1 = 0$ and for deformation parameter $1/\kappa=0.001$ as well as for initial value of total energy $E_{0,{\rm tot}} = 0.0015125$ lying down below the threshold $E_{0,{\rm th}} = 1/6$. Trajectory is partially chaotic.}\label{rysunek3}
\end{figure}

\eject

$~~~~~~~~~~~~~~~~~~$
\\
\\
\\
\\
\\
\begin{figure}[htp]
\includegraphics[width=\textwidth]{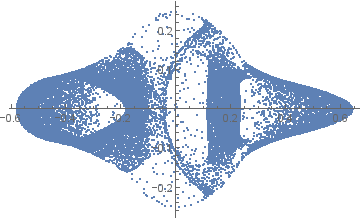}
\caption{Poincare map in two dimensional phase space $(x_2,p_2)$ for section $x_1 = 0$ and for deformation parameter $1/\kappa=0.0001$ as well as for initial value of total energy $E_{0,{\rm tot}} = 0.03125$ lying down below the threshold $E_{0,{\rm th}} = 1/6$. Trajectory is partially chaotic.}\label{rysunek3}
\end{figure}


\begin{thebibliography}{99}
\bibitem{lorenc}E.N. Lorenz, J. Atmos. Sci. 20, 130 (1963)
\bibitem{henon}M. Henon, C. Heiles, AJ. 69, 73 (1964)
\bibitem{rbsystem}S. Chandrasekhar, "Hydrodynamic and Hydromagnetic Stability" (Dover), ISBN 0-486-64071-X, 1982.
\bibitem{jerk}J.C. Sprott, Am J Phys. 65, 537 (1997)
\bibitem{duffing}G. Duffing, "Erzwungene Schwingungen bei Veränderlicher Eigenfrequenz", F. Vieweg u. Sohn, Braunschweig, 1918.
\bibitem{dpendulum}D.A. Wells, "Theory and Problems of Lagrangian Dynamics", New York: McGraw-Hill, pp. 13-14, 24, and 320-321, 1967.
\bibitem{dpendulum1}V.I. Arnold, "Problem in Mathematical Methods of Classical Mechanics", 2nd ed. New York: Springer-Verlag, p. 109, 1989.
\bibitem{quantumdamped}W.P. Schleich, "Quantum Optics in Phase Space", WILEY–VCH, Berlin, 2001.
\bibitem{tabor}M. Tabor, "Chaos and integrability in nonlinear dynamics",  New York: Wiley, 1989.
\bibitem{genhamhh}M.C. Gutzwiller, "Chaos in Classical and Quantum Mechanics", Springer-Verlag, Berlin, 1989.
\bibitem{henoncanondef}M. Daszkiewicz, Acta Phys. Polon. B 47, 2387 (2016); arXiv: 1610.08361 [physics.class-ph]
\bibitem{oeckl}R. Oeckl, J. Math. Phys. 40, 3588 (1999)
\bibitem{chi}M. Chaichian, P.P. Kulish, K. Nashijima, A. Tureanu, Phys. Lett. B
604, 98 (2004); hep-th/0408069
\bibitem{dasz1}M. Daszkiewicz,
Mod. Phys. Lett. A 23, 505 (2008); arXiv: 0801.1206 [hep-th]
\bibitem{class1}S. Zakrzewski, {"Poisson Structures on the Poincare
group"}; q-alg/9602001
\bibitem{class2}
Y. Brihaye, E. Kowalczyk, P. Maslanka, {"Poisson-Lie structure on Galilei
group"}; math/0006167
\bibitem{kappaP}J. Lukierski, A. Nowicki, H. Ruegg and V.N. Tolstoy, Phys. Lett.
B 264, 331 (1991)
\bibitem{kappaG}S. Giller, P. Kosinski, M. Majewski, P. Maslanka
and J. Kunz, Phys. Lett. B 286, 57 (1992)
\bibitem{lie1}
J. Lukierski and M. Woronowicz, Phys. Lett. B 633, 116 (2006); hep-th/0508083
\bibitem{qdef}O. Ogievetsky, W.B.  Schmidke, J. Wess, B. Zumino, Comm. Math. Phys.
150, 495 (1992)
\bibitem{paolo}
P. Aschieri, L. Castellani, A.M. Scarfone, Eur. Phys. J. C 7, 159
(1999); q-alg/9709032


\bibitem{mech}A. Deriglazov, JHEP 0303, 021 (2003); hep-th/0211105
\bibitem{ddddd}S. Ghosh, Phys. Lett. B 648, 262
(2007)
\bibitem{chaihydro1}M. Chaichian, M.M. Sheikh-Jabbari, A. Tureanu, Phys.
Rev. Lett. 86, 2716 (2001); hep-th/0010175
\bibitem{qmgnat}Kh.P. Gnatenko, V.M. Tkachuk, Phys. Lett. A 378, 3509 (2014); arXiv: 1407.6495 [quant-ph]
\bibitem{prefield}P. Kosinski, J. Lukierski, P. Maslanka, Phys. Rev.
D 62, 025004 (2000); hep-th/9902037
\bibitem{field}M. Chaichian, P. Pre\v{s}najder and  A. Tureanu,
Phys. Rev. Lett. 94, 151602 (2005); hep-th/0409096
\bibitem{fiorewess} G. Fiore, J. Wess, Phys. Rev. D
75, 105022 (2007); hep-th/0701078
\bibitem{2}S. Doplicher, K. Fredenhagen, J.E. Roberts, Phys. Lett. B 331, 39 (1994);
Comm. Math. Phys. 172, 187 (1995); hep-th/0303037
\bibitem{2a}A. Kempf and G. Mangano, Phys. Rev. D 55, 7909 (1997);
hep-th/9612084
\bibitem{recent}A. Connes, M.R. Douglas, A. Schwarz, JHEP 9802, 003 (1998); hep-th/9711162
\bibitem{string1}N. Seiberg and E. Witten, JHEP 9909, 032 (1999);
hep-th/9908142



\bibitem{drin}V.G. Drinfeld, Soviet Math. Dokl. 32, 254-258 (1985);
Algebra i Analiz (in Russian), 1, Fasc. 6, p. 114 (1989)
\bibitem{chaihydro2}M. Chaichian, M.M. Sheikh-Jabbari, A. Tureanu, Eur. Phys. J. C {36} (2004) 251; hep-th/0212259
\bibitem{exem}S. Gangopadhyay, A. Saha, A. Halder, Phys. Lett. A { 379} (2015) 2956; arXiv: 1412.3581 [hep-th]
\bibitem{romero1}
J.M. Romero, J.A. Santiago, J.D. Vergara, Phys. Lett. A 310, 9
(2003); hep-th/0211165
\bibitem{vv}Y. Miao, X. Wang, S. Yu, Annals Phys. 326, 2091 (2011); arXiv: 0911.5227 [math-ph]
\bibitem{kijanka}A. Kijanka, P. Kosinski, Phys. Rev. D 70, 12702 (2004); hep-th/0407246
\bibitem{p1}T. Bountis, H. Segur, F. Vivaldi, Phys. Rev. A 25, 1257 (1982)
\bibitem{p2}Y.F. Chang, M. Tabor, J. Weiss, J. Math. Phys. 23, 531 (1982)
\bibitem{p3}S. Wojciechowski, Phys. Lett. A 100, 277 (1984)
\bibitem{p10}A.P. Fordy, Physica D 52, 204 (1990)
\bibitem{bal1}A. Ballesteros, A. Blasco, Annals of Physics 325, 2787 (2010); arXiv: 1011.3005 [math-ph]
\bibitem{bal2}A. Ballesteros, A. Blasco, F.J. Herranz, J. Phys: Conf. Ser. 597 (2015) 012013; arXiv: 1503.09187 [nlin.SI]
\end{thebibliography}
\end{document}